# Influence of volatile solids and pH for the production of volatile fatty acids: batch fermentation tests using sewage sludge

**Dario Presti[1], Alida Cosenza[1], Fanny Claire Capri[2], Giuseppe Gallo[2], Rosa Alduina[2], Giorgio Mannina[1]***

[1]Dipartimento di Ingegneria, Università di Palermo, Viale delle Scienze, Ed. 8, 90128 Palermo, Italy

[2]Dipartimento di Scienze e Tecnologie Biologiche Chimiche e Farmaceutiche, Università di Palermo, Viale delle Scienze, Ed. 16, 90100 Palermo, Italy

- Corresponding author – e-mail: giorgio.mannina@unipa.it; tel: +3909123896556

**Abstract:**

The aim of this work was to study the effect of volatile suspended solid (VSS) and pH on volatile fatty acids (VFA) production from waste activated sludge (WAS) fermentation by means of batch tests. The final goal was to gain insights to enhance VFA stream quality, with the novelty of using WAS with high sludge retention time. Results revealed that the optimum conditions to maximize VFAs and minimize nutrients and non-VFA sCOD are a VSS concentration of 5.9 g/L and initial pH adjustment to pH 10. The WAS bacterial community structures were analysed according to Next Generation Sequencing (NGS) of 16S rDNA amplicons. The results revealed changes of bacterial phyla abundance in comparison with the batch test starting condition.

## 1 Introduction

The treatment and disposal of sewage sludge is nowadays a challenging issue to deal with during the wastewater treatment plants (WWTPs) operation. Indeed, an improper sewage sludge management/treatment may negatively affect the ecosystem's conditions, human health and may also contribute to climate change (Rorat et al., 2019). Moreover, sewage sludge management strongly affects the total operation costs of WWTPs accounting around 60% of the total costs (Maragkaki et al., 2018).

Since sewage sludge is an organic-rich material, the adoption of effective processing technologies for its treatment may have the advantage of minimizing its environmental impact while converting into recovered resources (Kehrein et al., 2020). Therefore, the adoption of proper technologies could contribute to save raw materials, preserve natural resources and sustain circular economy.

In recent decades, anaerobic digestion for biogas production has been widely investigated and implemented as sludge resource recovery technology (Demirbas et al., 2016; Kiselev et al., 2019). However, some major drawbacks still exist during the implementation/operation of this technology that weaken the advantages of anaerobic digestion in sewage sludge treatment (Luo et al., 2019). For example, the low biogas yield, the low amount of biogas produced and the low process carbon utilization rate occurring during the anaerobic digestion still require to be further investigated (Luo et al., 2019).

During the recent years, the acidogenic fermentation for volatile fatty acids (VFAs) production has been identified as a promising approach to use sewage sludge as a valuable resource (Liu et al., 2020). Indeed, by interrupting anaerobic digestion at the acidogenic stage, useful high value-added VFAs can be recovered, instead of low-value biogas.

VFAs are promising products since they have a wide potential of applications ranging from carbon source for biological nutrient removal process to the use as bioenergy resource for hydrogen and liquid biofuels generation (Kim et al., 2018). VFAs rich streams produced from sewage sludge fermentation can also be used as biopolymers precursor in the bioplastic industry, since they make a suitable feedstock for mixed microbial cultures (MMC) polyhydroxyalkanoates (PHA) production (Moretto et al., 2020). Some studies show a PHA content up to 59.2% of dry weight biomass when using VFAs obtained from sewage sludge acidification (Jia et al., 2014). In order to achieve process stability and high PHA yield it is crucial to manage the VFA rich streams quality to be used for PHA production. Indeed, high nutrients and non-VFA soluble chemical oxygen demand (sCOD) concentrations have adverse effect on the PHA production process (Tu et al., 2020; Zhang et al., 2019).

The current literature has deeply studied the influence of fermentation operational variables on the VFA production (Fang et al., 2020). Several literature studies have focused the attention on the optimization of operational variables to maximize VFA yield and productivity (Xu et al., 2021; Wang et al., 2021; Zhao et al., 2018; Yuan et al., 2016).

Wang et al. (2019) identified pH as key operational variable to improve the amount of VFA produced during WAS fermentation. Specifically, Wang et al. (2019) found that VFA production increases at high pH values. While Cokgor et al. (2006) perfomed fermentation tests at different VSS values, they found that VSS concentration strongly influences the VFA concentration in the fermented liquid (increases with VSS concentration). On the other hand, Yu et al. (2013) found that operating sewage sludge alkaline fermentation with high solid contents lead to increase in nutrient release during the process.

The aforementioned literature reveals that the knowledge on the influence of fermentation operational variables on the fermentation liquid quality (for a subsequent PHA production) is still lacking and controversial.

On the other hand, an increasing number of studies and techniques able to tune VFA rich stream quality for PHA production have been investigated (Zhang et al., 2018; Ye, Luo, et al., 2020; Tu et al., 2020; Tao et al., 2016). However, this kind of techniques such as struvite precipitation, ammonia stripping, addition of synthetic fatty acids substrate or membrane filtration, increase the process complexity and lead to additional operational costs making their applicability at full scale still unfeasible.

The aim of this study is to achieve an optimized set of operational variables for sewage sludge acidogenic fermentation. Specifically, pH and VSS were investigated as key operational variables during batch test operation. In particular, this study has a twofold aim: i. maximize the VFA yield of high SRT sludge; ii. obtain a high-quality VFA stream to maximize the PHA production, thus deleting the needs for further treatments and reducing the overall costs of PHA production process.

The study has the challenge of using sludge having high sludge retention time (SRT) as feedstock in view of a circular economy perspective (Mannina et al., 2021). Indeed, since high SRT reduces the sludge acidogenic potential, low SRT WAS has rarely been used in literature during acidogenic fermentation tests for VFA production.

## 2    Materials and methods

### 2.1    Wastewater treatment plant, sewage sludge and wastewater features

The sewage sludge used in this study was collected from the sludge recycle line of a real WWTP (i.e., Marineo) located at the north-western Sicilian coast (South of Italy). The WWTP design average daily flow was equal to 2,160 $m^3 d^{-1}$ (corresponding to 7,000 equivalent inhabitants).

The WWTP under study has a conventional activated sludge (CAS) scheme and operates at 35 days SRT. More precisely, the influent wastewater (WW) is first subjected to screening for solid separation and oil and grease removal and later the activated sludge processes are employed. The WWTP is also equipped with a bio-filtration system for water reuse.

During the experimental tests, wastewater was collected from the same plant and used when necessary to dilute the sewage sludge in view of adjusting the content of solids. Table 1 summarizes the features of the adopted sewage sludge and urban wastewater. The collected sewage sludge was used as both substrate and inoculum for the fermentation tests.

Table 1 Sewage sludge and wastewater features

| Parameters | Sewage sludge | Wastewater |
|---|---|---|
| pH | 7.1 | 7.3 |
| Total Suspended solids, TSS (g/L) | 14.1 | 0.44 |
| Volatile Suspended Solids, VSS (g/L) | 5.90 | 0.10 |
| Total Chemical Oxygen Demands, TCOD (g/L) | 13.39 | 0.54 |
| Soluble Chemical Oxygen Demands, sCOD (g/L) | 0.94 | 0.48 |
| Protein (g/L) | 0.67 | - |
| Carbohydrate (g/L) | 0.09 | - |
| Ammonium, $NH_4^+$-N (mg/L) | 8.3 | 23.5 |
| Phosphate, $PO_4^{3-}$-P (mg/L) | 20.8 | 9.5 |
| Volatile Fatty Acids, VFAs (mgCOD/L) | 60.2 | 21.3 |

## 2.2 Batch fermentation tests

Bench-scale batch fermentation tests have been performed in 1100 mL magnetic stirred glass bottles, equipped with two sampling ports for liquid and gas sampling and two electrode ports.

In Table 2 the batch tests details are reported. Specifically, 6 batch tests (T1 to T6) were performed in order to investigate the influence of each operational variable, namely VSS concentration and pH (Table 2).

**Table 2** Details of the performed batch fermentation tests

| Batch test | Details |
|---|---|
| T1 | VSS = 4 g/L <br> Uncontrolled pH |
| T2 | VSS = 5.9 g/L <br> Initial pH = 8 |
| T3 | VSS = 5.9 g/L <br> Uncontrolled pH |
| T4 | VSS= 5.9 g/L <br> Initial pH = 10 |
| T5 | VSS = 2.8 g/L <br> Uncontrolled pH |
| T6 | VSS = 5.9 g/L <br> pH = 10 (continuously adjusted) |

pH was adjusted by using 1N NaOH solution. The tests were conducted at room temperature (oscillating between 27 and 34 °C). According the typical duration for VFA production by anaerobic sludge fermentation reported in literature, each batch test lasted after 14 days operation (Yuan et al., 2011; Wang et al., 2019).

During the batch test, samples of the fermentation mixture and headspace gas were taken daily in order to analyse sCOD, VFA, ammonium ($NH_4^+$-N), phosphate ($PO_4^{3-}$-P) (around the VFA pick), $CO_2$ and $CH_4$ concentration. TSS, VSS and $PO_4^{3-}$-P were also measured during the initial and final test day, as well as the bacterial community structure analysis. While pH, temperature and redox potential have been continuously monitored by using a WiFi - Multi 3630 IDS "WTW and related probes.

## 2.3 Analytical methods

The fermentation mixture was withdrawn from each reactor and immediately filtered with 0.45 µm syringe filters for the composition analysis.

The analyses of sCOD, VSS, TSS, $NH_4^+$-N and $PO_4^{3-}$-P have been conducted according to the standard methods (APHA, 2005). In order to determine the VFAs concentration, 200 µL of the filtered fermentation mixture samples were transferred into gas chromatography (GC) vials and 800 µL of a stabilizing solution composed of $HgCl_2$ (0.5 g), phosphoric acid (5 mL, 100%) and hexanoic acid (0.54 mL, internal standard) were subsequently added. A gas chromatograph (GC) (Agilent Technologies 7820A), equipped with a flame ionization detector (FID) and a DB FFAA column (30 m x 0.25 x mm x 0.25 µm), was used for the detection of short chain fatty acids of up to 5 carbons (namely Acetic (HAc), Propionic (HPr), Isobutyric (HiB), Butyric (HB), Isovaleric (HiV) and Valeric (HV)). The samples were analysed by using the GC protocol proposed by Montiel-Jarillo et al. (2021). Carbon dioxide ($CO_2$) and Methane ($CH_4$) concentrations in the headspace gas were measured by using a GC (Thermo Scientific™ TRACE GC) equipped with an Electron Capture Detector (ECD).

Metagenomic analysis based on Next Generation Sequencing (NGS) was carried out on metagenomic DNA extracted according to Cinà et al (2019). DNA concentration was measured by reading absorbance at 260 nm with NanoDrop 2000c spectrophotometer (Thermo Fisher Scientific) and by running aliquots on 1% agarose gel with the addition of ethidium bromide to a final concentration of 0.5 μg/mL for UV visualization (data not shown). Metagenomic DNA was used as a template to amplify the region V3-V4 of the 16S rDNA gene using primers previously described (Takahashi et al., 2014). Amplification products were sequenced in one 300-bp paired-end run on an Illumina MiSeq platform (BMR Genomics, Padova, Italy). The raw 16S rDNA data were processed by using the QIIME2 environment (https://qiime2.org/) as paired-end sequences. In the denoising approach, overlapping paired-end reads were processed with the plug-in DADA2. Unique Amplicon Sequence Variants (ASVs) were assigned and aligned to the Greengenes reference database at 99% sequence similarity (https://greengenes.secondgenome.com/). For each sample, the number of ASVs and the percentages of relative abundances of phyla, orders, classes and families were determined.

NGS data have been registered with the BioProject database with the accession identifier PRJNA752593.

**2.4 Calculation of yield factors**

The COD solubilisation and VSS reduction have been calculated at the end of each batch test (Nazari et al., 2017).

Specifically, COD solubilisation has been calculated according to Equation 1 (Mohammad Mirsoleimani Azizi et al., 2021).

$$COD\ solubilization = \frac{sCOD_t - SCOD_0}{TCOD_0} \quad (1)$$

where subscripts (t) and (0) refer to the generic time and the initial time, respectively.

While, VSS reduction has been calculated according to Equation 2 (Mohammad Mirsoleimani Azizi et al., 2021).

$$VSS\ reduction = \frac{VSS_O - VSS_t}{VSS_0} \quad (2)$$

**3 Results and Discussion**

**3.1 Organic matter solubilization and VFA production**

sCOD patterns at different VSS concentration and pH values are shown in Figure 1a and 1b, respectively. It was observed that the hydrolysis in the tests without pH control (namely, T1, T3 and T5) was completed within 6 days as suggested by the sCOD peak concentration. This result is in line with a previous study by Yuan et al. (2011) that operated batch fermentation reactors under similar conditions.

In the tests with controlled pH (namely, T2, T4 and T6) the sCOD concentration peaked at different time depending on the pH control strategy used. For tests T2 and T4 where the pH was adjusted only at the beginning of the test, the peak occurred within the first 7 days, while for test T6, in which the pH was adjusted continuously, the sCOD concentration kept growing for the whole test period reaching the maximum of 4173 mgCOD/L at day 14. This behaviour is likely due to methanogenic microorganisms' inhibition. Indeed, methanogens are very sensitive to pH, their optimal range is around neutrality. For this reason the VFA produced in the uncontrolled or only initially controlled pH tests may be consumed by methanogens, but higher pH may hinder or prevent their transformation into $CH_4$, thereby stimulating VFA accumulation (Yu et al., 2013; Luo et al., 2019).

This hypothesis is justified by the pH evolution during the batch tests (Figure 1c). Indeed, during all tests (excepting for test T6) the pH rapidly reached a value of about 6.8 and remained stable.

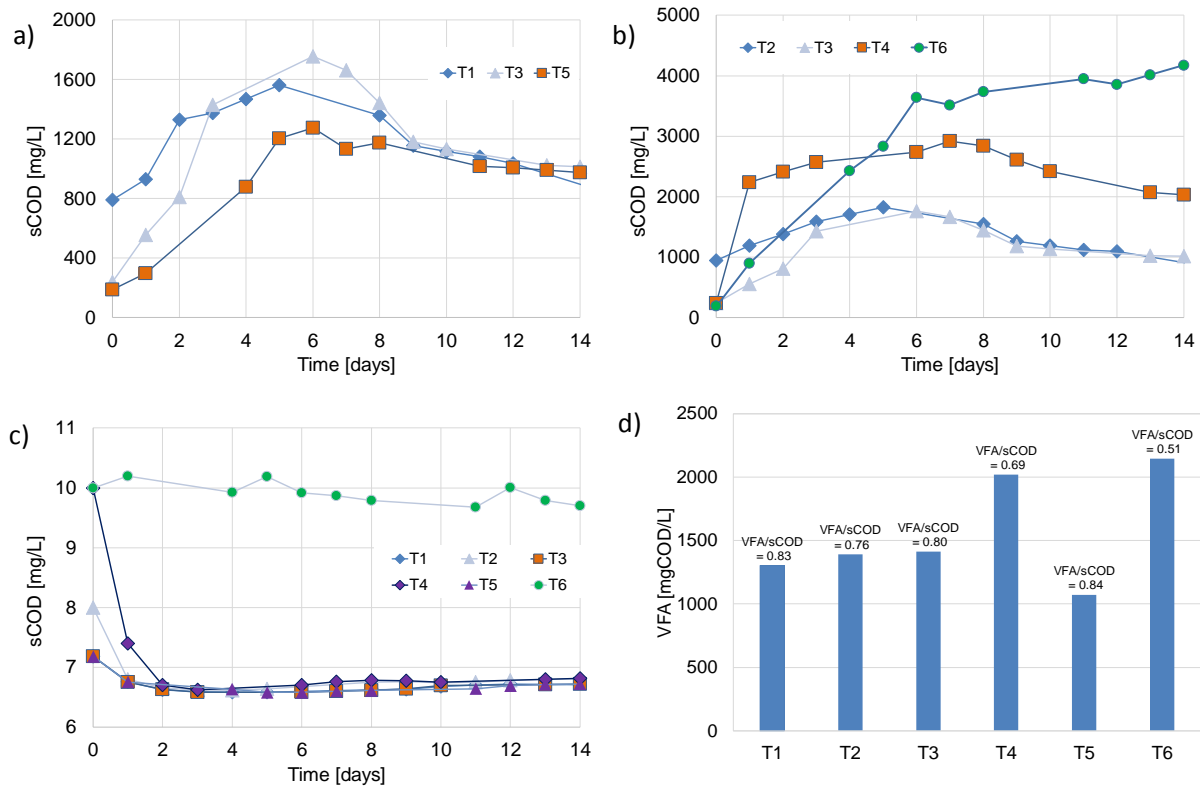

**Figure 1** - Influence of VSS concentration (a) and pH (b) on sCOD concentration; pH profiles (c) and maximum VFA concentration reached during each test (d).

The sCOD concentration obtained in this study at day 5 is lower than that reported in literature under similar conditions. For example, Yuan and co-authors obtained 2200 mg/L of sCOD concentration after 5 days of fermentation at a slightly lower temperature (24.6 °C) than that used in this study and without pH control (Yuan et al., 2011). This difference is likely due to the different SRT values of the initial sewage sludge. The SRT value reported by Yuan et al (2011) was 3 days, while here is 35 days. Indeed, it is commonly known that sludges with a lower SRT have an higher biodegradability and thus a better acidogenic potential (Xin et al., 2018; Montiel-Jarillo et al., 2021).

Figure 1d shows the VFA concentration obtained for each test at the day where maximum sCOD concentration was achieved. Among the six tests only T4 and T6 have provided the VFA concentration appropriate for a subsequent PHA production. Indeed, according to Bengtsson et al. (2017) suitable VFA feedstocks for teh subsequent PHA production should have a VFA concentration greater than 2gVFA/L.

Indeed, the highest absolute VFA concentration (2145 mgCOD/L) was reached during test T6, where pH was kept at 10 for the entire test duration. Despite the wide difference of the maximum sCOD concentration achieved during tests T6 and T4, only a slightly lower VFA concentration (2020 mgCOD/L) was obtained in test T4. The obtained VFA to sCOD ratio was equal to 0.69 and 0.51 for T4 and T6 tests, respectively (Figure 1d). Alkaline pH only promotes the hydrolysis and release of organic matter from the sludge particles into the liquid, while the neutral pH conditions that were established after less than 2 days in test T4 are conducive to the activity of acid-producing microorganisms (Ma et al., 2016). Further, literature suggests that low may hamper the PHA production process (Tu et al., 2020).

Therefore, since the final aim here is to product VFA rich stream suitable as feedstock for PHA production, T4 operating conditions can be considered the best result. Indeed, among the two tests with high VFA concentration (T4 and T6), test T4 has provided the VFA to sCOD ratio higher than that of T6.

Solubilization of sludge's particulate organic matter is attributed to the hydrolysis and disintegration of macromolecular biopolymers (proteins, lipids, and carbohydrates) to soluble monomers, that are used as substrate by acidogenic microorganism for VFA production, thus resulting in a reduction of the final VSS concentration. The VFA yields and VSS reduction percentages obtained in this study are reported in Figure 2.

The best VFA yield (383 mgCOD/gVSS) has been obtained at VSS concentration of 2.8 g/L (test T5), which is 60% higher than that obtained under the same conditions in test T3 (with a VSS concentration of 5.9 g/L) (Figure 2). A similar behaviour has been observed by Yuan et al. (2009) who have reported an increasing (46%) of VFA yield when VSS concentration decresed from 13 g/L to 4.8 g/L.

In terms of pH effect, the maximum value of VFA yield 363.5 mgCOD/gVSS has been achieved with the continuous pH 10 adjustment (test T6). The value obtained during test T4 is 5.8% lower than that of test T6.

In terms of VSS reduction, the highest VSS disintegration has been achieved in test T6 after 14 days followed by tests T4 and T2. This result reveals that alkaline conditions favour the organic matter solubilization and consequently the VSS reduction. Indeed, according to Yu et al. (2013), VSS reduction rate under alkaline conditions exceeded that of test T3 performed at the same VSS concentration but with uncontrolled pH.

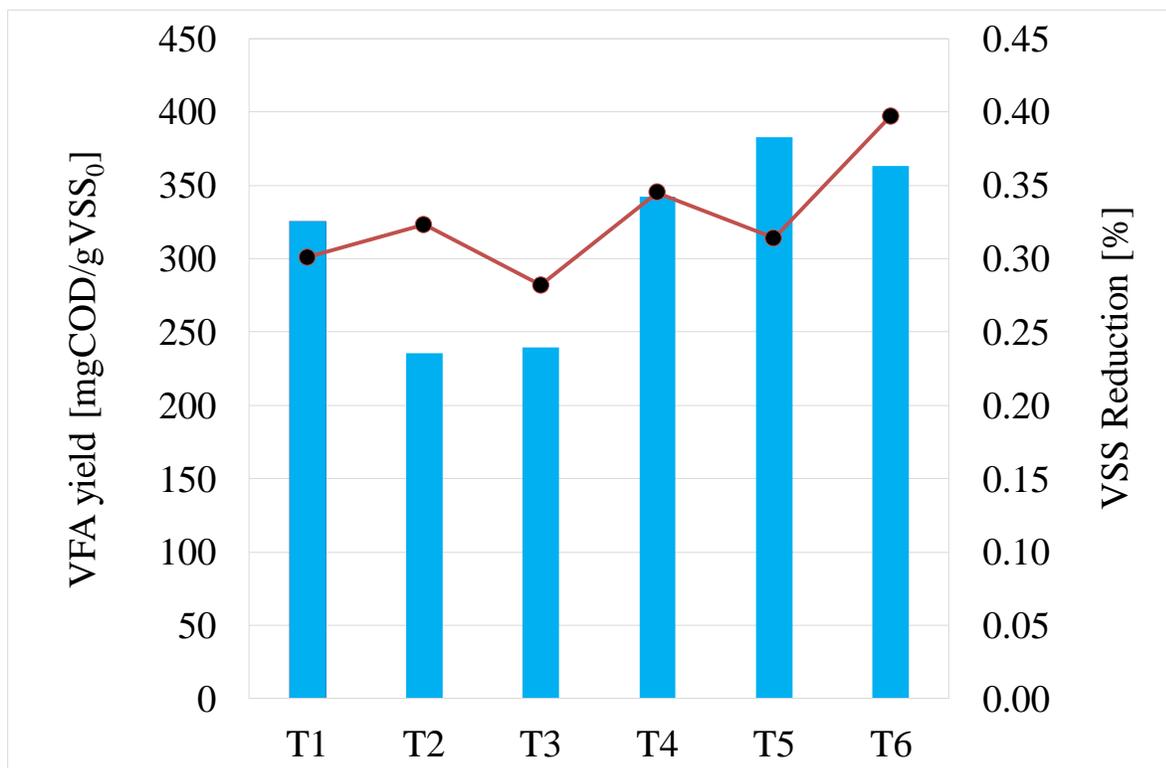

**Figure 2** – VFA yield (a) and VSS reduction percentage (b) for each batch test.

In terms of COD solubilization at the sCOD peak time the maximum value (19.3%) was reached during test T6 (Figure 3). Indeed, as discussed for VSS reduction percentage, alkaline conditions favour the organic matter solubilization. Nevertheless by comparing COD solubilization under uncontrolled pH (9.2%, test T3) and controlled pH (6.6%, test T2), this latter has achieved a lower COD solubilization (around 29% lower than T3). Regarding the influence of VSS concentration on COD solubilization it was observed that COD solubilization decreases with the increase of VSS concentration. Indeed, 9.2% and 12.6% of total COD was solubilized in test T3 and T5, respectively. These results are in line with the sientific literature, since it is reported that organic matter solubilization and sCOD yield usually decrease with the increase of solids concentration (Xiong et al., 2012).

For sake of completeness, the minimum, maximum and average $CO_2$ and $CH_4$ percentage in the batch reactors headspace off-gas are reported in Table 3.

**Table 3** Minimum, average and maximum values of $CO_2$ and $CH_4$ detected in reactors headspace during fermentation batch tests

|    | $CO_2$ | | | $CH_4$ | | |
|----|--------|--------|--------|--------|--------|--------|
|    | Min [%] | Average [%] | Max[%] | Min [%] | Average [%] | Max[%] |
| T1 | 9.73   | 12.97  | 18.19  | 8.43   | 26.51  | 51.26  |
| T2 | 10.83  | 15.26  | 24.11  | 8.22   | 25.76  | 58.13  |
| T3 | 1.93   | 9.93   | 17.86  | 2.45   | 15.90  | 28.46  |
| T4 | 0.89   | 7.76   | 15.94  | 0.21   | 25.54  | 51.94  |
| T5 | 3.81   | n/a    | n/a    | 0.56   | n/a    | n/a    |
| T6 | 0.42   | n/a    | n/a    | 0.56   | n/a    | n/a    |

Where n/a = data not availabel

Data reported in Table 3 suggest that the worse off-gas quality both in terms of $CO_2$ and $CH_4$ was obtained during test T2, thus corroborating the fact that the optmimal condition can be associated to test T4.

### 3.2 Ammonium and phosphorus release

Fermentation of sewage sludge results in the release of nitrogen and phosphorus from the decomposing biomass, mainly in $NH_4^+$-N and $PO_4^{3-}$-P form (Xiong et al., 2012; Wu et al., 2017). Figure 3a and b show the pattern of $NH_4^+$-N and $PO_4^{3-}$-P concentration measured during the tests, respectively. While, Figure 3c and d show the $NH_4^+$-N and $PO_4^{3-}$-P concentration, respectively, measured during the the sCOD peak for each tests. The $NH_4^+$-N concentration obtained during the sCOD peak for tests T1, T3 and T5 was respectively 40.0, 37.8 and 29.3 mg/L (Figure 3c). These results show that high VSS concentration increases the amount of $NH_4^+$-N concentration produced during the fermentation when the sCOD peak is reached. The $NH_4^+$-N concentration obtained during tests T2, T4 and T6 under the sCOD peak concentration was respectively 32.5, 49.2 and 28 mg/L (Figure 3c). Thus confirming literature results which suggest an increase of $NH_4^+$-N releasing with the pH increase (Yu et al., 2013; Ye, et al., 2020).

During test T6, $NH_4^+$-N concentration was 43% lower than that obtained during test T4. This result is likely due to the extreme pH that favors the ammoniacal nitrogen releasing mainly in form of free ammonia rather $NH_4^+$-N (Luo, et al., 2020).

The highest phospate realeasing (78 mg $PO_4^{3-}$-P/L) was achieved during test T3, while at lower VSS concentration $PO_4^{3-}$-P concentration was 50 and 29 mg $PO_4^{3-}$-P/L for test T1 and T5, respectively (Figure 3d). These results suggest that VSS concentration influence in the same way $NH_4^+$-N and $PO_4^{3-}$-P release.

Regarding the pH influence on $PO_4^{3-}$-P release opposite results compared to $NH_4^+$-N release have been here obtained. Indeed, $PO_4^{3-}$-P concentration obtained during tests T2, T4 and T6 was respectively 42, 36 and 27 mg $PO_4^{3-}$-P/L, thus suggesting that the highest $PO_4^{3-}$-P/L can be obtained under low pH values. Similar results were previously observed in literature by Wu and co-workers (Wu et al., 2017) who obtained acid conditions more favorable for $PO_4^{3-}$-P release.

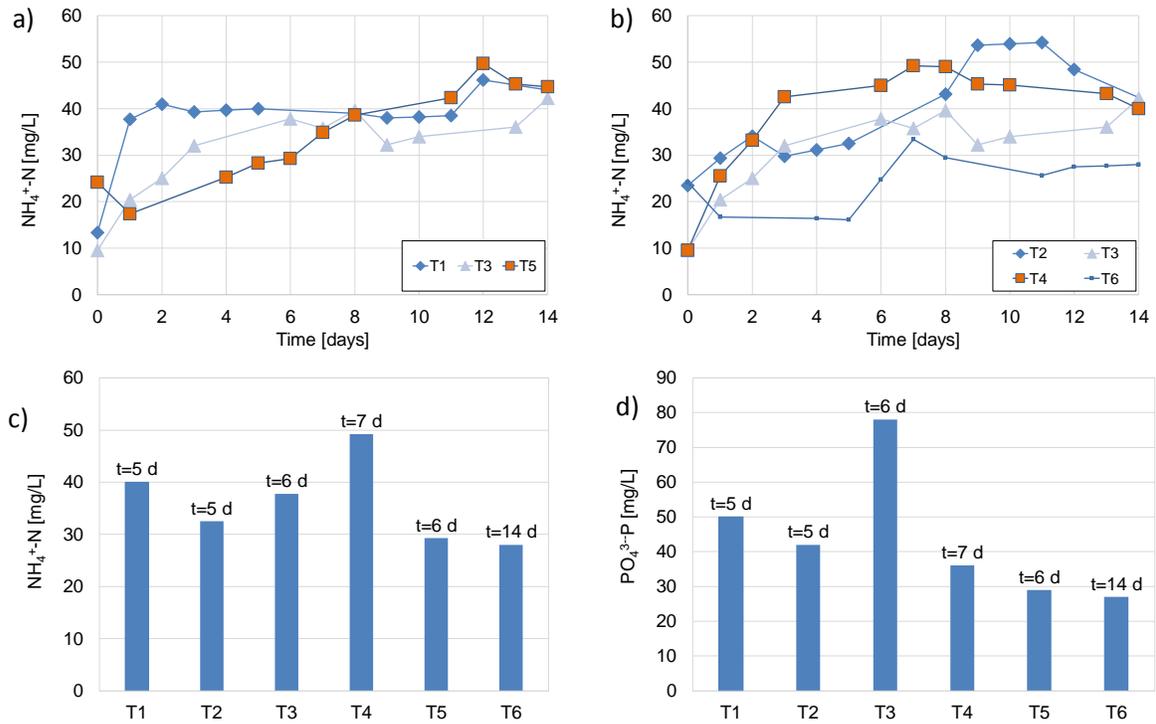

**Figure *3*** - Influence of VSS concentration (a) and pH (b) on ammonium release and ammonium (c) and phosphate (d) concentration at the sCOD peak day (t).

## 3.3 Bacterial community structures of activated sludge

The results of bacterial community and metagenomic DNA obtained for teh sludge used during tests T1 and T2 are here summarized. In particular, sewage collected at the endpoint of T1 and T2 was used as well as the sewage sludge collected from the sludge recycle line of the real WWTP used to inoculate T1 and T2, was used as starting condition (T0). Then, based on NGS analysis of the V3-V4 region of the 16S rRNA encoding gene amplified by PCR, the structure of the bacterial community was elucidated (Fig. 4). In particular, after denoising approach 388.230, 590.620 and 538.930 reads were obtained in T0, T1 and T2 samples, respectively. Based on ASV analysis, a total of 22 phyla were revealed apart unclassified bacteria (Table 4). In T0, T1 and T2 the most represented phyla are Proteobacteria and Bacteroidetes (Figure 4). However, in both T1 and T2 the Proteobacteria abundance decreased by about an half with compared to T0, with a dramatic increase of Firmicutes, Chloroflexi, Acidobacteria, Actinobacteria and Spirochaetes relative abundance (Figure. 4).

Similar results, highlighting the importance of Firmicutes for VFA production by activated sludges, were observed by Li et al. (2018) and Atasoy et al. (2019).

Interestingly, the activated sludge of T1 and T2 conditions showed similar bacterial community despite their different operating conditions.

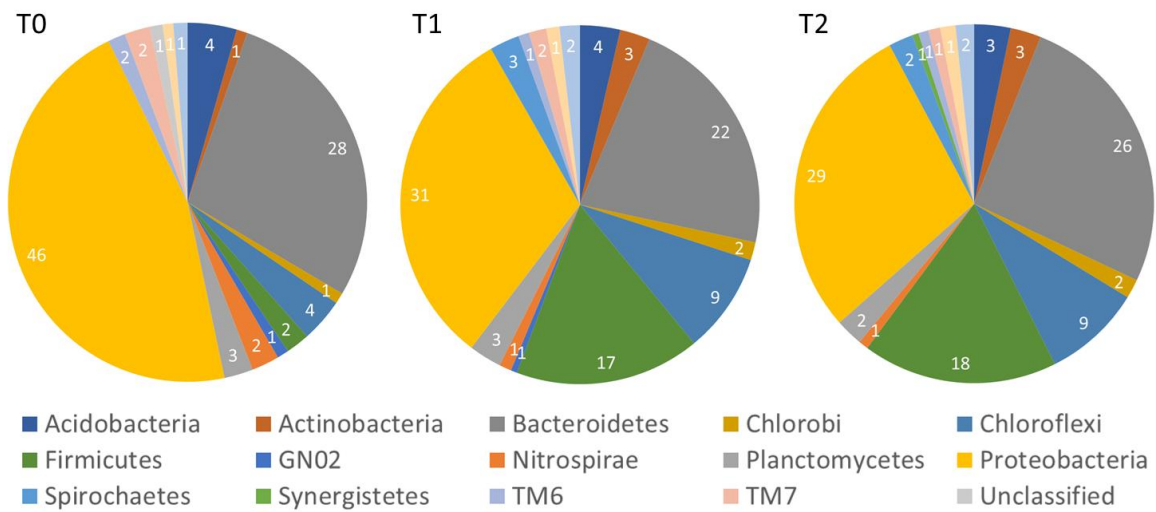

**Figure 4** - Relative abundance of bacterial phyla identified by NGS analysis of the V3-V4 region of the 16S rRNA encoding gene in the activated sludge of T0, T1 and T2 conditions, respectively. "Others" category includes phyla having

relative abundance lower than 0.5%. Numbers in the pies represent percentage value calculated on the total of each condition.

**Table 4** - Phyla revealed during T1 and T2 batch tests

| Phyla | Batch test | | |
|---|---|---|---|
| | T0 | T1 | T1 |
| Acidobacteria | 4.43 | 3.61 | 3.28 |
| Actinobacteria | 0.97 | 2.71 | 2.74 |
| Bacteroidetes | 28.10 | 22.05 | 25.99 |
| Caldiserica | 0.00 | 0.20 | 0.21 |
| Chlorobi | 0.99 | 1.61 | 1.72 |
| Chloroflexi | 3.94 | 9.05 | 8.86 |
| Firmicutes | 2.15 | 16.68 | 17.54 |
| Fusobacteria | 0.24 | 0.03 | 0.02 |
| Gemmatimonadetes | 0.22 | 0.20 | 0.17 |
| GN02 | 1.04 | 0.57 | 0.34 |
| GN04 | 0.35 | 0.13 | 0.11 |
| Nitrospirae | 2.49 | 1.10 | 0.87 |
| NKB19 | 0.00 | 0.37 | 0.37 |
| OD1 | 0.09 | 0.15 | 0.36 |
| Planctomycetes | 2.60 | 2.94 | 2.48 |
| Proteobacteria | 46.03 | 31.42 | 28.73 |
| Spirochaetes | 0.00 | 2.69 | 2.21 |
| Synergistetes | 0.02 | 0.44 | 0.56 |
| TM6 | 1.59 | 0.96 | 0.91 |
| TM7 | 2.28 | 1.58 | 1.07 |
| Unclassified | 1.17 | 0.04 | 0.00 |
| Verrucomicrobia | 0.95 | 1.19 | 1.38 |
| WS3 | 0.35 | 0.25 | 0.07 |

## 4  Conclusions

Sewage sludge collected from full scale WWTP was here adopted for batch fermentation tests investigating different conditions in terms of VSS concentration and pH. The main aim was to test the technical feasibility of obtaining high-quality VFA stream by using high SRT sludge as inoculum.

On the basis of the results obtained here, the optimum conditions to maximize VFAs and minimize nutrients and non-VFA sCOD are at VSS concentration of 5.9 g/L and initial pH adjustment to pH 10. NGS analysis has demonstrated changes in the bacterial community of inoculum sludge compared during the batch tests.

**Acknowledgments**

This work was funded by the project "Achieving wider uptake of water-smart solutions—WIDER UPTAKE" (grant agreement number: 869283) financed by the European Union's Horizon 2020 Research and Innovation Programme, in which the last author of this paper, Giorgio Mannina, is the principal investigator for the University of Palermo. The Unipa project website can be found at: https://wideruptake.unipa.it/. The authors wish to thank Dr. Ylenia Di Leto for technical support.

29  Yu, H., Wang, Z., Wang, Q., Wu, Z., and Ma, J., 2013, Disintegration and acidification of MBR sludge under alkaline conditions: *Chemical Engineering Journal*, v. 231, p. 206–213.

30  Yuan, Y., Liu, Y., Li, B., Wang, B., Wang, S., and Peng, Y., 2016, Short-chain fatty acids production and microbial community in sludge alkaline fermentation: Long-term effect of temperature: *Bioresource Technology*, v. 211, p. 685–690.

31  Yuan, Q., Sparling, R., and Oleszkiewicz, J.A., 2011, VFA generation from waste activated sludge: Effect of temperature and mixing: *Chemosphere*, v. 82, no. 4, p. 603–607.

32  Yuan, Q., Sparling, R., and Oleszkiewicz, J.A., 2009, Waste activated sludge fermentation: Effect of solids retention time and biomass concentration: *Water Research*, v. 43, no. 20, p. 5180–5186.

33  Zhang, D., Jiang, H., Chang, J., Sun, J., Tu, W., and Wang, H., 2019, Effect of thermal hydrolysis pretreatment on volatile fatty acids production in sludge acidification and subsequent polyhydroxyalkanoates production: *Bioresource Technology*, v. 279, no. January, p. 92–100.

34  Zhang, L., Liu, He, Zheng, Z., Ma, H., Yang, M., and Liu, Hongbo, 2018, Continuous liquid fermentation of pretreated waste activated sludge for high rate volatile fatty acids production and online nutrients recovery: *Bioresource Technology*, v. 249, no. August 2017, p. 962–968.
26